\documentclass[epjST,final]{svjour}[]
\usepackage{amsmath,amsfonts,amssymb,lmodern}

\usepackage[usenames,dvipsnames]{color}

\usepackage{enumerate}

\usepackage{cite}

\usepackage[rightcaption]{sidecap}

\makeatletter
\newenvironment{HorizFig}{\SC@float[c]{figure}}{\endSC@float} %caption zentr./seitl.
\makeatother

\renewcommand{\d}{\mbox{d}}

\newcommand{\mean}[1]{\left < #1 \right>}
\newcommand{\abs}[1]{\left | #1 \right |}

\usepackage[pdftex]{graphicx}

\begin{document}
%
% \title{Isotropic \& Anisotropic Self-Propelled Motion}
%  \subtitle{Insights from a geometric approach}
\title{A geometric approach to self-propelled motion in isotropic \& anisotropic environments}
\author{Robert Gro\ss{}mann\inst{1}\fnmsep\thanks{\email{grossmann@physik.hu-berlin.de}} \and Fernando Peruani\inst{2}
\and Markus B\"ar\inst{1} }
\institute{Physikalisch-Technische Bundesanstalt, Abbestr. 2-12, 10587 Berlin, Germany \and Laboratoire J. A.
Dieudonn\'{e}, Universit\'{e} de Nice Sophia Antipolis, UMR 7351 CNRS, Parc Valrose,
F-06108 Nice Cedex 02, France}
\abstract{
We propose a geometric perspective to describe the motion of self-propelled particles moving at constant
speed in $d$ dimensions.  
We exploit the fact that the vector that conveys the direction of motion of the particle performs a
random walk on a $(d-1)$-dimensional manifold. 
We show that the particle performs isotropic diffusion in $d$-dimensions if the manifold corresponds to a hypersphere. 
In contrast, we find that the self-propelled particle exhibits anisotropic diffusion if this manifold corresponds to a deformed 
hypersphere
(e.g. an ellipsoid). 
This simple approach provides an unified framework to deal with isotropic as well as anisotropic diffusion of particles
moving at constant speed in any dimension. 
} %end of abstract
\maketitle
\section{Introduction}
\label{sec:intro}

We find examples of self-propelled entities in a remarkably large variety of chemical~\cite{paxton2004, mano2005,
rucker2007, howse2007, golestanian2007, golestanian2009}, physical~\cite{kudrolli2008, deseigne2010, weber2013} and
biological~\cite{vicsek2012} systems. 
The non-equilibrium nature of these active systems has recently been investigated through the 
development of what has been called active soft matter~\cite{vicsek2012,marchetti2012}. 
In active systems, such as self-propelled particle (SPP) systems, the constant conversion of energy into work --
used by the particles to self-propel in a dissipative medium -- drives the system out of thermodynamic
equilibrium leading to remarkable physical properties for both, interacting as well as non-interacting SPP systems. 
In interacting SPP systems, large-scale collective motion patterns~\cite{vicsek2012,marchetti2012,grossmann2014} and
non-equilibrium clustering~\cite{peruani2006,peruani2012,peruani2013} are observed in the presence of a
velocity-alignment interaction for homogeneous media with periodic boundary conditions. 
Moreover, the presence of fluctuations in both, the moving direction of the particle
and its speed -- typically related to fluctuations of the self-propelling engine -- leads to bistability
of macroscopic order and disorder \cite{grossmann2012}. For non-interacting active particles, these
non-thermal fluctuations result in complex non-equilibrium transients in the mean squared
displacement~\cite{peruani2007}
and anomalous (non-Maxwellian) velocity distributions~\cite{golestanian2009, romanczuk2011}.
Interestingly, the physics of active systems is remarkably different when other boundary conditions
are used~\cite{denis} -- the lack of momentum conservation in active systems induces non-classical particle-wall
interactions which allow, for instance, the rectification of particle motion~\cite{galajda2007, wensink2008, wan2008,
tailleur2009, radtke2012}.
Besides, non-interacting SPPs can exhibit spontaneous particle trapping and subdiffusion in heterogeneous
media~\cite{chepizhko2013a,chepizhko2013b}. 

In this work, we focus on the simplest class of SPP systems that we can think of: non-interacting SPPs moving at
constant speed. 
Despite their simplicity, these SPP models find applications in artificial active particles 
such as vibration-driven rods and disks~\cite{kudrolli2008, deseigne2010, weber2013}, light-driven~\cite{jiang2010,
golestanian2012, theurkauff2012, palacci2013} and chemically-driven~\cite{paxton2004, mano2005, rucker2007, howse2007,
golestanian2007, golestanian2009} SPPs at low density as well as diluted bacterial systems~\cite{dorota}. 
Here, we contribute to the theoretical description developed in previous works~\cite{mikhailov_self_1997,
peruani2007, golestanian2009, romanczuk2011} by proposing a geometric perspective onto self-propelled motion. 
A geometric formulation of the model has several advantages. First, geometric insights yield the possibility of 
understanding observations intuitively and allow to generalize
two dimensional SPP models to any spatial dimension immediately. Furthermore, we show that the same theoretical framework
enables us to describe SPP motion in anisotropic environments as occurs, for instance, in experiments with eukaryotic
cells moving on a pre-patterned surfaces~\cite{Csucs_locomotion_2007,ziebert_effects_2013}. 
Moreover, the geometric formulation simplifies the mathematical modeling itself: we derive stochastic equations of motion from a 
geometric principle which is reflected by a Fokker-Planck equation defining an unique stochastic
process. Accordingly, ambiguities related to the interpretation of Langevin equations \cite{gardiner_stochastic_2009} do
not occur. 
In short, we propose a simple unified framework to model isotropic as well as anisotropic
diffusion of SPPs in any dimension. 

This paper is organized as follows. In section \ref{sec:GeomView}, we present our geometric
approach to model self-propelled motion and discuss technical details in
section \ref{sec:GenMod}. 
Based on this general framework, we review isotropic self-propelled motion in section \ref{sec:isoSPP} focusing on two
spatial dimensions and subsequently generalizing to arbitrary dimensions. 
In section \ref{sec:AnIsoSpp}, we use the geometric approach to study SPPs in anisotropic environments. We summarize
and discuss our results in section \ref{sec:Summary}.

\section{A geometric view on self-propelled motion}
\label{sec:GeomView}

We consider non-interacting SPPs moving at constant speed $v_0$. 
The moving direction is denoted by a vector $\vec{e}(t)$, called \textit{director} for short in
the following. 
The dynamics of an active particle is then given by the following differential equation:
\begin{align} \label{eq:first}
 \frac{\d \vec{r}(t)}{\d t} = \vec{v}(t) = v_0 \hspace{0.03cm} \hat{\vec{e}}(t), \quad \hat{\vec{e}}(t) =
\vec{e}(t)/\!\hspace{0.01cm}\abs{\vec{e}(t)} \!. 
\end{align}
The spatial dynamics is solved by
\begin{align} \label{eq:firstSol}
 \vec{r}(t) = \vec{r}(t_0) + v_0 \! \int_{t_0}^t  \! \mbox{d} t' \, \frac{\vec{e}(t')}{\abs{\vec{e}(t')}} .
\end{align}
In this sense, the spatial dynamics $\vec{r}(t)$ of the SPP is subordinated to the 
stochastic dynamics of the director $\vec{e}(t)$. Once the stochastic properties of the director $\vec{e}(t)$ are known,
the motion of a SPP in space can be characterized using Eq. \eqref{eq:firstSol}. 
For example, the correlation function of the director and the mean squared displacement of a SPP are linked by
the Taylor-Kubo formula
\cite{Taylor_diffusion_1922,kubo_statistical_1957,ebeling_statistical_2005}: 
 \begin{align}
   \label{eqn:gen:Kubo:rel}
  \mean{\left | \vec{r}(t) - \vec{r}(t_0)\right |^2} = \mean{\left | \Delta \vec{r}(t) \right |^2 } = v_0^2
\int_{t_0}^t \d t' \int_{t_0}^{t} \d t ''\, \mean{\hat{\vec{e}}(t') \cdot \hat{\vec{e}}(t'')} \!.
 \end{align}
Having understood that the motion of an active particle moving at constant speed is prescribed by the dynamics of the
director, we focus on the latter henceforward. We will show that a geometric view onto the director dynamics simplifies
the modeling of self-propelled motion in arbitrary dimensions, both in isotropic and anisotropic environments.

\begin{figure}[tb]
 \begin{center}
   \includegraphics[width=0.8\textwidth]{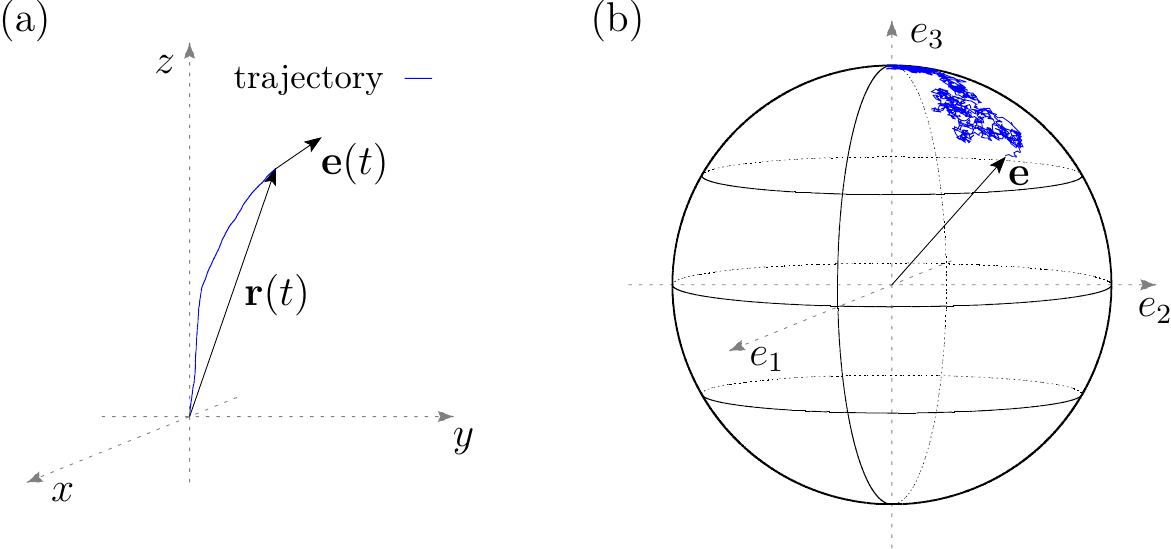}
 \end{center}
 \caption{(a) Visualization of the trajectory $\vec{r}(t)$ of a SPP in three dimensions and
(b) the corresponding diffusive motion of the moving direction $\vec{e}(t)$ on a sphere $S^2$.  
Notice that the SPP moves in $d$ dimensions -- as shown 
in panel (a) -- and the moving direction $\vec{e}(t)$ lives 
on a surface of dimension $d-1$, panel (b).  
}
 \label{fig:iso2D} 
\end{figure}

Let us briefly recall how self-propelled motion is modeled in two
dimensions~\cite{schienbein_langevin_1993,mikhailov_self_1997}. Usually, the director is parametrized by an angle
$\varphi(t)$ as follows
 \begin{align}
  \label{eqn:2d:1:r:2}
  \vec{e}(t) = \begin{pmatrix}
                \cos \varphi(t) \\ \sin \varphi(t)
               \end{pmatrix} \! , 
 \end{align}
where the polar angle obeys the stochastic differential equation
 \begin{align}
  \label{eqn:2d:1:r:3}
   \frac{\mbox{d} \varphi(t)}{\mbox{d} t} = \sqrt{2 D}\,  \xi(t). 
 \end{align}
Hence, the trajectory described by the director $\vec{e}(t)$ is equivalent to the trajectory of a Brownian particle
moving on a circle. Accordingly, the two dimensional motion of a SPP is subordinated to the ordinary
Brownian motion of the director $\vec{e}(t)$ on a manifold that in this case is simply a  circle. This geometric interpretation of Eq.
\eqref{eqn:2d:1:r:2}-\eqref{eqn:2d:1:r:3} enables us to generalize the model to arbitrary dimensions as follows. The
motion of a SPP in three dimensions is controlled by the director dynamics $\vec{e}(t)$ which describes 
diffusive motion on a sphere as illustrated in Fig.~\ref{fig:iso2D}. Thus, the director describes Brownian motion on a
$(d-1)$-sphere in the case of $d$-dimensional self-propelled motion. 
In other words, the description of isotropic
self-propelled objects is directly related to random walks on
hyperspheres \cite{brillinger_particle_1997} where Eq. \eqref{eq:firstSol} links these two processes. 

If the trajectory $\vec{e}(t)$ corresponds to Brownian motion on a perfectly isotropic sphere, the 
resulting
trajectory $\vec{r}(t)$ of the SPP is isotropic in space as well. Accordingly, the SPP does not possess a preferential
direction of motion. Following the geometric picture drawn above, the description of anisotropic self-propelled motion
is straightforward: the director $\vec{e}(t)$ describes Brownian motion on a compact manifold different from a hypersphere. Such 
a manifold could be, for instance, an ellipsoid. 

In summary, the SPP motion is isotropic if the director performs Brownian motion on a hypersphere. In contrast, SPP motion is
(typically) anisotropic if this manifold corresponds to the surface of a ``deformed'' hypersphere (the details on the properties 
of the manifold are given in section \ref{sec:GenMod}). 

Thus, to close our simple SPP model we have to (i) specify the surface on which the director moves and (ii) provide the
dynamics for the director on this surface. 
We argued above that the motion of the director on the manifold is purely diffusive. Hence, the dynamics of the
probability
density for the director pointing in a certain direction, $P(\vec{e},t)$, is determined by the diffusion equation
\begin{align}
 \label{eqn:MG:DiffEqn}
 \frac{\partial P(\vec{e},t)}{\partial t} = D \Delta_{\hspace{0.01cm}\vec{e}} P(\vec{e},t). 
\end{align}
The operator $\Delta_{\vec{e}}$ denotes the Laplace-Beltrami operator on the particular manifold. The parameter
$D$ measures the strength of the fluctuations acting on the director which defines the diffusion
coefficient on the manifold but not the diffusion coefficient of the SPP. 
Notice that according to
\eqref{eqn:MG:DiffEqn}, $D$ determines the characteristic time scale of relaxation but, by definition,
does not affect stationary states. 
Depending on the particular system under consideration, the noise strength $D$ may depend
on additional parameters, for example the speed of the particle describing speed dependent
persistence~\cite{mikhailov_self_1997, romanczuk2011}. In this study, we will treat it as an independent parameter. 

\subsubsection*{The spatial diffusion coefficient}

Let us suppose that the underlying stochastic process described by the director is stationary
\cite{gardiner_stochastic_2009} in the sense that observables do not depend on the instant of time at which they have
been
measured. 
Under this assumption, the director correlation function does depend on time differences only and we may write $C(\Delta t) =
\mean{\hat{\vec{e}}(t') \cdot \hat{\vec{e}}(t'+\Delta t)}$. We define the correlation time $\tau_{c}$ as integral of the
correlation function\footnote{This definition is valid only if the correlation function does not oscillate. Moreover,
we implicitly assume that the integral over the correlation function converges excluding anomalous diffusion
processes in this context. }: 
\begin{align}
 \label{eqn:def:corr:time:8}
 \tau_{c} = \int_0^{\infty} \d \Delta t\, C(\Delta t). 
\end{align}
Using this definition and assuming $\langle \vec{e}(t\to\infty) \rangle=0$, one can rewrite integral
\eqref{eqn:gen:Kubo:rel} in order to obtain the long time limit 
\begin{align*}
 \mean{\left | \Delta \vec{r}(t) \right |^2} \sim 2 \tau_{c} v_0^2 t = 2 d \mathcal{D}_{x} t. 
\end{align*}
The last expression in the equation above is the defining relation of the spatial diffusion coefficient
$\mathcal{D}_{x}$ which is related to the correlation time and the spatial dimensionality via 
\begin{align}
 \label{eqn:gen:diff:coeff:kobu:21}
 \mathcal{D}_x = \frac{\tau_{c} v_0^2}{d}\, .
\end{align}
The preceding discussion illustrates that the stochastic motion on curved space described by the director as well as
the properties of SPPs are inextricably linked. 

\section{Langevin dynamics of the director}
\label{sec:GenMod}

According to the model explained in the previous section, the dynamics of the director corresponds to Brownian
motion on a compact manifold~\cite{brillinger_particle_1997,zinn_quantum_2002}. 
We assume that this compact manifold can be embedded in $d$ dimensions
and parametrize the points of this space by Cartesian coordinates: $\vec{e}=(e_1,e_2,...,e_d)$. 
In subsequent sections, we will associate the direction of motion of the SPP with a point on the manifold via $\hat{\vec{e}}(t) 
= \vec{e}(t)/\!\hspace{0.01cm}\abs{\vec{e}(t)}$, cf. Eq.~\eqref{eq:first}.
% 
% Notice that implicitly we are assuming that every point on the manifold has associated a director (i.e. a moving direction). 
%
% This means that if we know the position of the on the manifold, we know the moving direction of the particle. 
%
% For simplicity, we assume that the director is given by the vector that goes from the origin to the given position on the manifold, as indicated in Fig.~\ref{fig:iso2D}(b), 
% that we normalize as $\hat{\vec{e}}(t) =
% \vec{e}(t)/\!\hspace{0.01cm}\abs{\vec{e}(t)}$, Eq.~\eqref{eq:first}.  
% 
In this section, we derive the stochastic differential equations that govern the stochastic motion of the director $\vec{e}$ on 
the manifold whose dynamics is defined by the
Fokker-Planck equation \eqref{eqn:MG:DiffEqn}.

At a more mathematical level, the comment above means that the director is a $d$-dimensional vector $\vec{e} \in \mathbb{R}^d$, 
while its motion is
confined to a surface of a geometrical object (a compact manifold) which is embedded in $d$ dimensions. 
Let the points on the manifold be parametrized by $n<d$ generalized coordinates $\varphi_\mu$, where $\mu=1,2,...,n$. 
For concreteness, imagine the coordinates $\varphi_\mu$ to be the angles which parametrize the points on an unit sphere.
The transformation of coordinates from the Cartesian laboratory reference frame to generalized
coordinates reads
\begin{align}
 \label{eqn:var:trans}
    e_i &= f_i(\varphi_1,\varphi_2,...,\varphi_n),
\end{align}
where $i=1,2,...,d$. We define the Jacobian matrix $J$ of this transformation as
 \begin{align}
   J_{i\mu} = \frac{\partial e_i}{\partial \varphi_\mu}. 
 \end{align}
The central objects for the description of the diffusion on a manifold are the metric tensor $g$ whose elements read
 \begin{align}
   g_{\mu \nu} = \sum_{i=1}^{d} J_{i\mu} J_{i \nu}, 
 \end{align}
as well as the inverse metric tensor $\Lambda = g^{-1}$. 

In \eqref{eqn:var:trans}, we expressed the director by generalized coordinates. The Laplacian operator in
Eq.~\eqref{eqn:MG:DiffEqn} must be
expressed by these coordinates as well \cite{jost_riemannian_2008}
\begin{align}
 \Delta_{\hspace{0.01cm} \vec{e}} = \frac{1}{\sqrt{\abs{g}}} \sum_{\mu=1}^n \sum_{\nu=1}^n \partial_{\mu} \left ( \!
\sqrt{\abs{g}} \, \Lambda_{\mu \nu}
\partial_{\nu} \right )\! ,
\end{align}
where $\partial_\mu$ is a shorthand for the derivative with respect to $\varphi_\mu$ and we denote
the determinant of the metric tensor by $\abs{g}$. The probability
density $P(\vec{e},t)$ becomes a function of the generalized variables and it is normalized
according to the condition
 \begin{align}
   \label{eqn:NormCond1}
   \int \! \d \varphi_1 \int \! \d \varphi_2 \dots \int \! \d \varphi_n  \, \sqrt{\abs{g}} \, P(\vec{e},t) = 1. 
 \end{align}
We introduce the probability density $p \hspace{0.035cm}\! \left (\hspace{-0.02cm}\left \{ \varphi \right \}\!,t \right
)$ to find the director pointing
in a certain direction by absorbing the measure factor in \eqref{eqn:NormCond1} as 
 \begin{align}
   p \hspace{0.035cm}\! \left (\hspace{-0.02cm}\left \{ \varphi \right \}\!,t \right
) = \sqrt{\abs{g}} \, P(\vec{e},t). 
 \end{align}
The dynamics of $p \hspace{0.035cm}\! \left (\hspace{-0.02cm}\left \{ \varphi \right \}\!,t \right)$ follows from its
definition and the general form
of the Laplacian operator on the manifold. We reorganize terms in the resulting equation such that it takes the
standard form of a Fokker-Planck equation:
\begin{align}
 \label{eqn:FPE:gen2}
 \frac{\partial \hspace{0.035cm}p \hspace{0.035cm}\! \left (\hspace{-0.02cm}\left \{ \varphi \right \}\!,t
\right)}{\partial t} = D \sum_{\mu,\nu} \partial_{\mu}\partial_\nu \! \left ( 
\Lambda_{\mu \nu} p \right ) - D \sum_{\mu,\nu} \partial_\mu \left [ p \left ( \partial_\nu \Lambda_{\mu \nu} +
\Lambda_{\mu \nu} \partial_\nu \log
\! \sqrt{\abs{g}} \, \right ) \right ] \! .
\end{align}
From this Fokker-Planck equation which is written in the so called \textit{Ito} form \cite{gardiner_stochastic_2009}
we can immediately read off the corresponding Langevin dynamics for the generalized variables. The first term encodes
the strength of the fluctuations (diffusion term), whereas the second term describes a
deterministic drift. In order to write down the Langevin dynamics for the generalized coordinates, we introduce the
generalized noise amplitudes $\sigma$, a $(n\times n)$-matrix having the property $\Lambda = \sigma \cdot
\sigma^T$. Hence, the stochastic differential equation for the generalized coordinates in Ito (I) interpretation reads 
 \begin{align}
  \mbox{(I)} \quad \frac{\d \varphi_\mu}{\d t} &= \sqrt{2 D} \sum_{\alpha} \sigma_{\mu \alpha} \, \xi_\alpha(t) +
D\sum_{\nu} \left (
\partial_\nu \Lambda_{\mu \nu} + \Lambda_{\mu \nu} \partial_\nu \log \! \sqrt{\abs{g}} \, \right ) \! .
\label{eqn:GenLangMod:Ito}
 \end{align}  
The corresponding differential equation in Stratonovich (S) interpretation is obtained by applying the usual rules of stochastic calculus \cite{gardiner_stochastic_2009}:
 \begin{align}
  \mbox{(S)} \quad \frac{\d \varphi_\mu}{\d t} &= \sqrt{2 D} \sum_{\alpha} \sigma_{\mu \alpha} \, \xi_\alpha(t) +  D
\sum_{\nu} \left
(\partial_\nu \Lambda_{\mu \nu} - \sum_{\alpha} \sigma_{\nu \alpha} \partial_\nu \sigma_{\mu \alpha} + \Lambda_{\mu \nu}
\partial_\nu
\log \! \sqrt{\abs{g}} \, \right )\!.   \label{eqn:GenLangMod:Strato}
 \end{align}  
 The random
processes $\xi_\alpha(t)$ denote independent Gaussian random processes with zero mean and temporal
$\delta$-correlations. Since
the dynamics involves multiplicative noise -- the noise amplitude depends on the state
of the system -- a word on the interpretation of the Langevin equation is in order. Both Langevin equations,
\eqref{eqn:GenLangMod:Ito} and \eqref{eqn:GenLangMod:Strato}, describe the same physics since
they were derived from the same Fokker-Planck equation. In other words, one can choose the interpretation
depending on which equation is easier to handle analytically or more convenient to use in a numerical experiment. 

The equations \eqref{eqn:GenLangMod:Ito} and \eqref{eqn:GenLangMod:Strato} describe diffusion on a compact manifold
which is embedded in $d$-dimensional
space. They are simplified considerably if the manifold is parametrized using an orthogonal basis. 
In this
particular case, the metric tensor is diagonal and we find
 \begin{align}
   \label{eqn:diag:met:21}
   g_{\mu \nu} &= g_{\mu\mu} \delta_{\mu\nu} , \qquad \Lambda_{\mu \nu} = \frac{\delta_{\mu\nu}}{g_{\mu\mu}}, \qquad
\sigma_{\mu\nu} = \frac{\delta_{\mu\nu}}{\sqrt{g_{\mu\mu}}}
 \end{align}  
and thus
\begin{subequations}
 \label{eqn:GenLangMod:diagMet}
 \begin{align}
  \mbox{(I)} \quad \frac{\d \varphi_\mu}{\d t} &= \sqrt{\frac{2 D}{g_{\mu\mu}}} \, \xi_\mu(t) + D \left ( \partial_\mu
g_{\mu\mu}^{-1} + g_{\mu\mu}^{-1} \, \partial_\mu \log \sqrt{\abs{g}} \, \right ) \! ,\\
  \mbox{(S)} \quad \frac{\d \varphi_\mu}{\d t} &= \sqrt{\frac{2 D}{g_{\mu\mu}}} \, \xi_\mu(t) + \frac{D}{2} \left (
\partial_\mu g_{\mu\mu}^{-1} + g_{\mu\mu}^{-1} \, \partial_\mu \log \abs{g} \right ) \!. \label{eqn:GenLangMod:diagMetb}
 \end{align}  
\end{subequations}

In \eqref{eqn:GenLangMod:diagMet}, both drift and diffusion terms are due to \textit{geometric} properties of the
manifold and proportional to the strength of the fluctuations $D$. 
This last point, i.e. that drift and diffusion terms are proportional to $D$, distinguishes our approach 
from the motion of particles subjected to an external forcing. 
In contrast to motion induced by an external force, the drift and diffusion terms are not independent, in the sense
that by taking the limit $D\to0$, both terms vanish, 
while we expect the drift term to survive in the presence of an external forcing. 
In summary, the strength of
an external field or the response of a particle to an external stimulus requires some additional parameter (force
strength). 
For anisotropic environments, we will make use of a parameter characterizing the anisotropy of the environment
for instance, but this will not be equivalent to an external force.

In general, the Langevin equations \eqref{eqn:GenLangMod:Ito} and \eqref{eqn:GenLangMod:Strato} determine the temporal evolution 
of the generalized coordinates given any arbitrary metric
tensor. 
However, we want to interpret the vector $\vec{e}$ as the
direction of motion of a self-propelled particle, cf. Eq. \eqref{eq:first} and Eq. \eqref{eqn:var:trans}.  
Therefore, the embedding of the manifold is not arbitrary. 
In principle, the vector $\vec{e}$ could be shifted by a constant vector $\vec{e}_0$ (independent of the
generalized coordinates) without changing the metric tensor or the equations of motion. However, the velocity of
the SPP would shift according
to Eq.~\eqref{eq:first}: $\vec{v}= v_0 \hat{\vec{e}} \rightarrow v_0 \left (\vec{e} + \vec{e}_0\right )\!/\!\abs{\vec{e} + 
\vec{e}_0}$. The constant shift would
imply ballistic motion into one particular direction determined by $\vec{e}_0$. We fix this constant by requiring that 
 \begin{align}
\label{eq:manifold}
\int \! \d \varphi_1 \int \! \d \varphi_2 \dots \int \! \d \varphi_n  \, \sqrt{\abs{g}} \, \vec{e}\! \left (
\left \{ \vec{\varphi} \right \} \!,t\right ) = 0. 
 \end{align}
Geometrically speaking, this condition is fulfilled if the manifold is embedded in such a way that the center of mass of the 
surface parametrized by $\vec{e}$
coincides with the origin. For example, the sphere in Fig.~\ref{fig:iso2D}b is centered. 

The framework above will be illustrated by several examples in the subsequent chapters. 
% We will restrict our discussion to simple cases where the equations \eqref{eqn:GenLangMod:diagMet} are sufficient for the description of the system.
According to our previous discussion, we follow the recipe as outlined before:
 \begin{enumerate}[(i)]
  \item Parametrization of the manifold by a convenient set of generalized coordinates;
  \item Calculation of the Jacobian as well as the metric tensor;
  \item Derivation of the inverse metric tensor and noise amplitudes;
  \item Deduction of the Fokker-Planck and Langevin equation from the general expressions discussed in this section;
  \item Solution of the model either analytically using Fokker-Planck equations or numerically via Lan\-ge\-vin
simulations. 
 \end{enumerate}

\section{Isotropic self-propelled motion}
\label{sec:isoSPP}

The director $\vec{e}(t)$ parametrizes points on a hypersphere $S^{d-1}$ for self-propelled motion in $d$ spatial dimensions. We 
 choose $\abs{\vec{e}(t)}
= 1$ for convenience. In this section, we start out with the case $d=2$, derive the equations of motion in three spatial
dimensions and then conclude with a general discussion for higher dimensional cases. 

\subsection{Self-propelled motion in d=2}

In this subsection, we discuss the motion of a SPP in $d=2$ which was studied in
\cite{schienbein_langevin_1993,mikhailov_self_1997}. In two spatial dimensions, the director which
determines the direction of motion of a particle describes diffusion on a circle of unit radius. Hence, the
direction of motion can be parametrized by a single angle $\varphi$ according to Eq. \eqref{eqn:2d:1:r:2}. 
The Langevin dynamics of $\varphi$ is well-known, cf. Eq. \eqref{eqn:2d:1:r:3}. 
As an exercise, we will deduce this expression using  our general approach discussed in the previous section. Since the
director
is described by a single variable, Greek indices can be omitted in this context. From Eq.~\eqref{eqn:2d:1:r:2}, we
obtain the Jacobian
 \begin{align*}
   J = 
     \begin{pmatrix}
       \frac{\partial \cos \varphi }{\partial \varphi} \\ \frac{\partial \sin \varphi}{\partial \varphi}
     \end{pmatrix} 
     = 
     \begin{pmatrix}
       -\sin \varphi \\ \;\cos \varphi 
     \end{pmatrix} 
 \end{align*}
and the corresponding metric tensor
 \begin{align*}
   g = J^T \cdot J = 1,
 \end{align*}
which is just a number in this case. Hence, we obtain expression~\eqref{eqn:2d:1:r:3} by inserting the metric tensor
in the
general expression \eqref{eqn:GenLangMod:diagMet}. The corresponding Fokker-Planck equation is obtained from
\eqref{eqn:FPE:gen2} by inserting the metric, or directly from \eqref{eqn:2d:1:r:3}:
 \begin{align}
 \partial_t p(\varphi,t) = D \partial^2_{\varphi} \, p(\varphi,t) .
 \end{align}
The solution of this Fokker-Planck equation is obtained from separation of variables
 \begin{align} \label{eq:sol_Diff_eq}
  p(\varphi,t|\varphi_0,t_0) = \frac{1}{2\pi} + \frac{1}{\pi} \sum_{m=1}^{\infty} \cos [m(\varphi - \varphi_0)] e^{-m^2
D (t-t_0)}, 
 \end{align}
where $t \ge t_0$. The correlation function of the director reads 
 \begin{align}
  \mean{\vec{e}(t) \cdot \vec{e}(t')} &= \int_{0}^{2 \pi} \d \varphi \, \cos \! \left ( \varphi - \varphi' \right
) p(\varphi,t| \varphi',t') 
  = e^{-D \abs{t-t'}} \, .   \label{eqn:corr:func:time:2d}
 \end{align}
From the Taylor-Kubo relation \eqref{eqn:gen:Kubo:rel}, we obtain the correlation time as well as the diffusion
coefficient 
 \begin{align}
   \tau_{c} = \frac{1}{D}\,  , \qquad \mathcal{D}_x = \frac{v_0^2}{2 D} \, ,
 \end{align}
which were derived in~\cite{schienbein_langevin_1993,mikhailov_self_1997} first. The correlation time determines the
crossover timescale at which the persistent ballistic motion of a particle turns into random diffusive motion. 
 
\subsection{Self-propelled motion in d=3}

We follow the recipe described above in order to derive the equations of motion for a SPP in three
dimensions which is less trivial than the previous example as shown in the following. In three spatial dimensions, the
director is parametrized by two angles $\theta$ and $\varphi$ via
 \begin{align}
  \vec{e} =
  	\begin{pmatrix}
                e_1 \\ e_2 \\ e_3
        \end{pmatrix}
  =  
	\begin{pmatrix}
                \sin \theta \cos \varphi \\ \sin \theta \sin \varphi \\ \cos \theta
        \end{pmatrix}\!,
 \end{align}
where $\varphi \in [0,2\pi)$ and $\theta \in [0,\pi)$ in our convention. At first, we calculate the Jacobian
 \begin{align*}
   J = \begin{pmatrix}
          \frac{\partial e_1}{\partial \theta}   &\; &  \frac{\partial e_1}{\partial \varphi}  \\
          \frac{\partial e_2}{\partial \theta}   &\; &  \frac{\partial e_2}{\partial \varphi}  \\
          \frac{\partial e_3}{\partial \theta}   &\; &  \frac{\partial e_3}{\partial \varphi}  \\
       \end{pmatrix}
       = 
       \begin{pmatrix}
                \cos \theta \cos \varphi &\; & - \sin \theta \sin \varphi \\
                \cos \theta \sin \varphi &\; &   \sin \theta \cos \varphi \\
                -\sin \theta             &\; &  0 
        \end{pmatrix}
 \end{align*}
as well as the metric tensor
 \begin{align}
  g = J^T \cdot J =  
     \begin{pmatrix}
           1 &\; & 0 \\ 0 &\; & \sin^2 \theta
     \end{pmatrix} \!. 
 \end{align}
The metric tensor has a diagonal form, i.e. we can use \eqref{eqn:GenLangMod:diagMetb} in order to derive the Langevin
dynamics of the angles:   
\begin{subequations}
 \label{eqn:GenLangMod:diagMet3d}
 \begin{align}
  \mbox{(S)} \,\quad \frac{\d \theta}{\d t}  &= \sqrt{2 D} \, \xi_\theta(t) + \frac{D}{\tan \theta} , \\
  \mbox{(S)} \quad \frac{\d \varphi}{\d t} &= \frac{\sqrt{2 D}}{\sin \theta} \, \xi_\varphi(t) . 
 \end{align}  
\end{subequations}
The corresponding Fokker-Planck equation reads
 \begin{align}
  \frac{\partial p(\theta,\varphi,t)}{\partial t} = - \frac{\partial }{\partial \theta} \left [ \frac{D p }{\tan \theta}
\right ] +  D \frac{\partial^2 p}{\partial \theta^2} +  \frac{D}{\sin^2 \theta} \frac{\partial^2 p}{\partial
\varphi^2}. 
 \end{align}
By integration over the polar angle $\varphi$, we obtain a Fokker-Planck equation for the marginal probability density
$\tilde{p}(\theta,t)$
 \begin{align}
  \frac{\partial \tilde{p}(\theta,t)}{\partial t} = - \frac{\partial }{\partial \theta} \left [ \frac{D \tilde{p} }{\tan
\theta} \right ] +  D \frac{\partial^2 \tilde{p}}{\partial \theta^2} 
  \end{align}
with its time-dependent solution
 \begin{align}
   \label{eqn:MPDF:Dens:3d:8}
   \tilde{p}(\theta,t|\theta_0,t_0) = \frac{\sin \theta}{2} \sum_{m=0}^\infty (2m+1) P_m(\cos{\theta})
P_m(\cos{\theta_0}) e^{-m (m+1) D (t-t_0) }. 
 \end{align}
We denote Legendre polynomials by $P_m$. Equation \eqref{eqn:MPDF:Dens:3d:8} is the probability density function to find
a particle moving with a angle $\theta$ versus the $e_3$-axis given that it started moving into the direction
$\theta_0$ at $t = t_0$. 
% 
% Since the underlying dynamics is Markovian, we can calculate the correlation function from
% this distribution: 
% 
Since the motion is isotropic and the dynamics is Markovian, we can always assume that the initial direction of motion equals the 
direction $\vec{e}=(0,0,1)$ without loss of generality and use Eq. \eqref{eqn:MPDF:Dens:3d:8} to compute the correlation 
function: 
\begin{align}
  \mean{\vec{e}(t) \cdot \vec{e}(t')} = \int_{0}^{\pi} \d \theta \, \cos \theta \, \tilde{p}(\theta,t|0,t') = e^{-2
D \abs{t-t'}}. 
 \end{align}
Note, that the correlation function is exponentially decreasing as well, i.e. the behavior of a SPP in
$d=3$ is similar to the motion in $d=2$. However, the correlation time in two dimensions is a factor of two larger than
in three dimensions. Accordingly, the diffusion coefficient is reduced by a factor of three: 
 \begin{align}
      \tau_{c} = \frac{1}{2 D}\,  , \qquad \mathcal{D}_x = \frac{v_0^2}{6 D}\, .
 \end{align}
 
\subsection{Self-propelled motion in arbitrary dimensions}
\label{sec:SPP_dDim}

Surprisingly, the properties of SPPs in two and three spatial dimensions are qualitatively
very similar. Their characteristics like the diffusion coefficient and velocity correlation time differ by prefactors.
Naturally, the question arises whether the behavior of SPPs is qualitatively similar in all
spatial dimensions or whether a crossover/critical dimension exists above which the qualitative behavior changes. In
this section, we answer this question by showing that the velocity correlation function of a SPP is
indeed exponentially decreasing irrespective of the spatial dimensionality. Furthermore, we use this opportunity to
present another approach for the description of the diffusion of the director. 

The parametrization of the director by angles as discussed in previous sections becomes increasingly difficult in
higher dimensions. For completeness, we present the stochastic dynamics of the angles in appendix \ref{sec:App:DDDDD}. 
Here, we apply a different approach to study the dynamics of the director in $d$ dimensions using Cartesian
coordinates \cite{brillinger_particle_1997}: 
\begin{subequations}
 \label{eqn:iso:diff:alt}
 \begin{align}
  \mbox{(I)} \quad \frac{\d e_i}{\d t} &= - D (d-1) e_i + \sqrt{2D} \, \sum_{j=1}^d \left ( \delta_{ij} - e_i e_j \right
) \xi_j(t)
\label{eqn:iso:diff:alt:b} , \\
  \mbox{(S)} \quad \frac{\d e_i}{\d t} &= \sqrt{2D} \, \sum_{j=1}^d \left ( \delta_{ij} - e_i e_j \right ) \xi_j(t).
\label{eqn:iso:diff:alt:c}
 \end{align}    
\end{subequations}
This parametrization is completely equivalent to the parametrization by angles which can be proven by explicitly
carrying out the change of variables from $\left \{ e_i \right \}$ to angles $\left \{ \varphi_{\mu} \right \}$.
However, care must be taken that the rules for the transformation of variables from ordinary calculus can only be
applied if the stochastic differential equation is interpreted in the sense of Stratonovich
\cite{gardiner_stochastic_2009}, i.e. in Eq. \eqref{eqn:iso:diff:alt:c}. 
In this context, we stress again the fact that both stochastic differential equations in \eqref{eqn:iso:diff:alt}
describe the same stochastic process. 
In particular, the length of the director $\abs{ \hspace{-0.001cm} \vec{e}(t)} = 1$ is conserved in both
cases\footnote{In particular, the term $\dot{e}_i \propto - D (d-1) e_i$ in Eq. \eqref{eqn:iso:diff:alt:b} ensures the
conservation of the length $\abs{\vec{e}} = 1$ if the corresponding equation is interpreted in Ito sense. }. They do
only appear different at a first glance because their interpretation is different. 
Moreover, the dynamics described by \eqref{eqn:iso:diff:alt} is Markovian. 

In the following, we will calculate the correlation function of the director in arbitrary dimensions. For this purpose,
we multiply \eqref{eqn:iso:diff:alt:b} by $\vec{e}(t')$, where $t' \le t$, and average over many realization of the
process. Therefore, we obtain a linear equation for the director correlation function
 \begin{align*}
  \frac{\d \!\mean{\vec{e}(t) \cdot \vec{e}(t')} }{\d t} &= - D (d-1) \mean{\vec{e}(t) \cdot \vec{e}(t')}. 
 \end{align*}    
This equation is readily solved with the appropriate initial condition. We obtain: 
 \begin{align*}
  \mean{\vec{e}(t) \cdot \vec{e}(t')} &= e^{-D(d-1) \abs{t-t'}}. 
 \end{align*}    
Hence, the correlation function decreases exponentially in \textit{all} spatial dimensions $d$. We obtain the
correlation time by integration, cf. Eq. \eqref{eqn:def:corr:time:8}, 
 \begin{align}
   \tau_{c} = \frac{1}{D(d-1)}
 \end{align}    
in this general case and the diffusion coefficient of a SPP in $d$ dimensions:
 \begin{align} 
  \mathcal{D}_x = \frac{v_0^2}{D d (d-1)}. 
 \end{align}    
Consistently, these general results reproduce our earlier results. Thus, we have shown that the qualitative properties
of isotropic self-propelled motion do not depend on the spatial dimensionality. The only difference is the dependence of
the correlation time on the spatial dimension. 

\section{Anisotropic self-propelled motion}
\label{sec:AnIsoSpp}

We aim at generalizing the isotropic SPP model to account for anisotropic environments. 
We stress that anisotropic diffusion is fundamentally different from a biased motion due to some external
forcing which trivially induces biased particle flow and ballistic
motion at large timescales rather than diffusion. Here, we focus on SPPs with an anisotropic directional
persistence which may be due to an anisotropic substrate. For the sake of
clarity, we restrict our discussion to the
two-dimensional case. However, the generalization to $d=3$ is straightforward from the general framework discussed in
previous sections.

Let us assume that the director moves on a manifold parametrized in polar coordinates in the following way\footnote{The choice 
of this parametrization a priori excludes
a number of curves in two dimensions which
are not relevant in the context of self-propelled motion. } 
\begin{align}
 \label{eqn:dist:circ:2d}
 \vec{e} = \begin{pmatrix}
             e_x \\ e_y
           \end{pmatrix}
 = r(\varphi) 
                      \begin{pmatrix}
                        \cos \varphi \\ \sin \varphi
                      \end{pmatrix} \! ,\quad
 \hat{\vec{e}} =      \begin{pmatrix}
                        \cos \varphi \\ \sin \varphi
                      \end{pmatrix} \!,
\end{align}
where $r(\varphi)$ is a $\pi$-periodic function\footnote{If $r(\varphi)$ is not $\pi$-periodic, the \textit{center of
mass} of the curve would be shifted away from zero. Consequently, this would imply biased ballistic motion. By assuming
$\pi$-periodicity, equation \eqref{eq:manifold} is obeyed. }. Curves parametrized in this way are \textit{distorted
circles}, e.g.  an ellipse as illustrated in Fig.
\ref{fig:ellipse}. The ellipse is parametrized by
 \begin{align*}
    r(\varphi) = \frac{1}{\sqrt{1-\epsilon^2 \cos^2 \varphi}} ,
 \end{align*}
where the parameter $\epsilon \in [0,1)$ denotes the eccentricity. The limiting case of a circle
(isotropic motion) corresponds to  $\epsilon = 0$ and  $r(\varphi) = 1$. 

The angle $\varphi$ determines the direction of motion of the SPP in two
dimensions. The derivation of the dynamics of $\varphi$ is straightforward from the general discussion in section
\ref{sec:GenMod}.
We only need to calculate the metric tensor which reads:
\begin{align}
  g(\varphi) = \left [ r(\varphi) \right ]^{2} + \left [ \frac{\d r(\varphi)}{\d \varphi} \right ]^{\!2}\! .
\end{align}
The concrete form of the dependence $r(\varphi)$ is of minor importance, since the Langevin and Fokker-Planck
dynamics for the angle $\varphi$ and the probability density $p(\varphi,t)$, respectively, depend on $g(\varphi)$ only. 
 \begin{HorizFig}[50][tb]
    \includegraphics[width=0.4\textwidth]{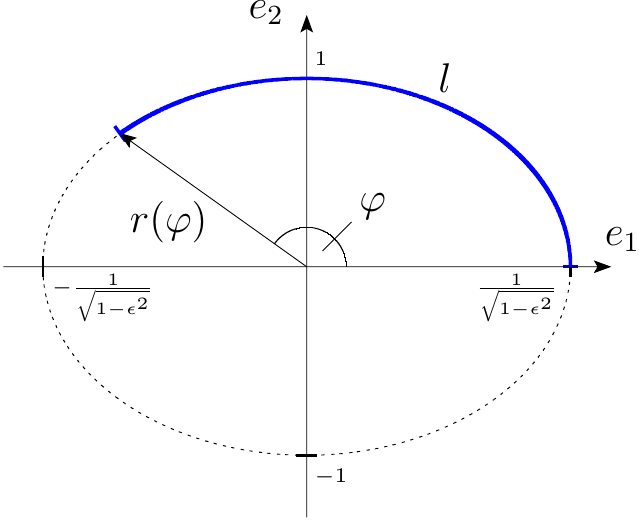}
 \caption{Visualization of the director parametrized by \eqref{eqn:dist:circ:2d}. Exemplarily,
we have chosen an ellipse with eccentricity $\epsilon$. Furthermore, we represent the arc length $l$ by a blue line, cf.
\eqref{eqn:ArcLength:Def}. }
 \label{fig:ellipse}
 \end{HorizFig} 

We derive the Langevin equation for the angle $\varphi$ from \eqref{eqn:GenLangMod:diagMet}: 
\begin{align}
   \label{eqn:Langevin:Aniso:2d}
   \mbox{(S)} \quad \frac{\d \varphi}{\d t} &= \sqrt{\frac{2D}{g(\varphi)}} \, \xi(t). 
\end{align}
We conclude that the anisotropy is reflected by a state dependent noise amplitude (\textit{anisotropic directional
persistence}). If the particle moves in a
direction $\varphi$ where $g(\varphi)$ is large (preferred direction of motion), the level of fluctuations is reduced.
In contrast, fluctuations are amplified in regions where $g(\varphi)$ is small. Note, however, that the particle moves
at constant speed. The anisotropic motion arises from the non-homogeneous distribution of  times a particle moves
in a certain spatial direction. 

The corresponding Fokker-Planck equation for the dynamics of the probability density $p(\varphi,t)$ is obtained from
the Langevin equation \eqref{eqn:Langevin:Aniso:2d} for the angle: 
\begin{align}
 \label{eqn:FPE:2d:ANI}
 \frac{\partial p(\varphi,t)}{\partial t} = D \frac{\partial}{\partial \varphi} \left [
\frac{1}{\sqrt{g(\varphi)}} \frac{\partial}{\partial \varphi} \left ( \frac{p(\varphi,t)}{\sqrt{g(\varphi)}} \right )
\right ] \! .
\end{align}
Without further knowledge, this equation is difficult to solve. However, our geometric interpretation of the problem
allows us to derive the time dependent solution of the Fokker-Planck equation without specification of $g(\varphi)$. 
In order to do so, we exploit the fact that our problem can be mapped to the
one-dimensional diffusion equation which can be easily solved analytically. 
In a first step, we parametrize the manifold in terms of the the arc length
\cite{bronshtein_handbook_2007} 
\begin{align}
 \label{eqn:ArcLength:Def}
 l = \int_0^\varphi \d \varphi' \, \sqrt{g(\varphi')} \, ,
\end{align}
cf. Fig. \ref{fig:ellipse} for an visualization, giving this change of variables a transparent interpretation. We denote
the total length of the curve by
\begin{align*}
 L = \int_0^{2\pi} \d \varphi' \, \sqrt{g(\varphi')} \, .
\end{align*}
We introduce the probability density $\tilde{p}(l,t)$ to find a particle within the length interval $[l,l+\d l]$. By
substitution of variables in \eqref{eqn:FPE:2d:ANI}, we indeed find -- as expected -- the ordinary diffusion equation
\begin{align*}
 \frac{\partial \tilde{p}(l,t)}{\partial t} = D \, \frac{\partial^2 \tilde{p}(l,t)}{\partial l^2} . 
\end{align*}
The time dependent solution of the diffusion equation in one dimension with periodic boundary conditions,
$\tilde{p}(l+L,t) = \tilde{p}(l,t)$, is found by separation of variables. In fact, it has already been discussed in the
context of isotropic self-propelled motion in two dimensions. Using this solution and subsequent resubstitution of
$\varphi$ for the arc length yields the time dependent solution of Eq. \eqref{eqn:FPE:2d:ANI}: 
\begin{align} 
 \label{eqn:gen:sol:2d:P:ani:11}
 p(\varphi,t| \varphi_0,t_0) = \frac{\sqrt{g(\varphi)}}{L} \, \left \{ 1+2 \sum_{m=1}^{\infty} \cos \left [
\frac{2 \pi m}{L} \int_{\varphi_0}^\varphi \d \varphi' \, \sqrt{g(\varphi')}\right ] e^{- \left
(2 \pi m/L \right )^{2} D \, (t-t_0) } \right \} \! .
\end{align}
Notice that $p(\varphi,t| \varphi_0,t_0)\neq p(\varphi_0,t| \varphi,t_0)$ in contrast to what we learned for isotropic diffusion, c.f. \eqref{eq:sol_Diff_eq}. The stationary distribution $p_0(\varphi)= p(\varphi,t| \varphi_0,t_0 \rightarrow - \infty)$ is given by geometric
properties only  
\begin{align}
 \label{eqn:stat:prob:ani:phi}
 p_0(\varphi) = \frac{\sqrt{g(\varphi)}}{L} ,
\end{align}
since the strength of angular fluctuations $D$ determines the timescale of relaxation, but it does not influence the
stationary distribution as argued in the context of equation \eqref{eqn:MG:DiffEqn}. The distribution of the arc length
$\tilde{p}_0(l)$ is a constant,
\begin{align}
 \label{eqn:stat:prob:ani:l}
 \tilde{p}_0(l) = \frac{1}{L},
\end{align}
solely determined by the total arc length $L$. 
\begin{figure}[tb]
\begin{center}
 	\includegraphics[width=0.6\textwidth]{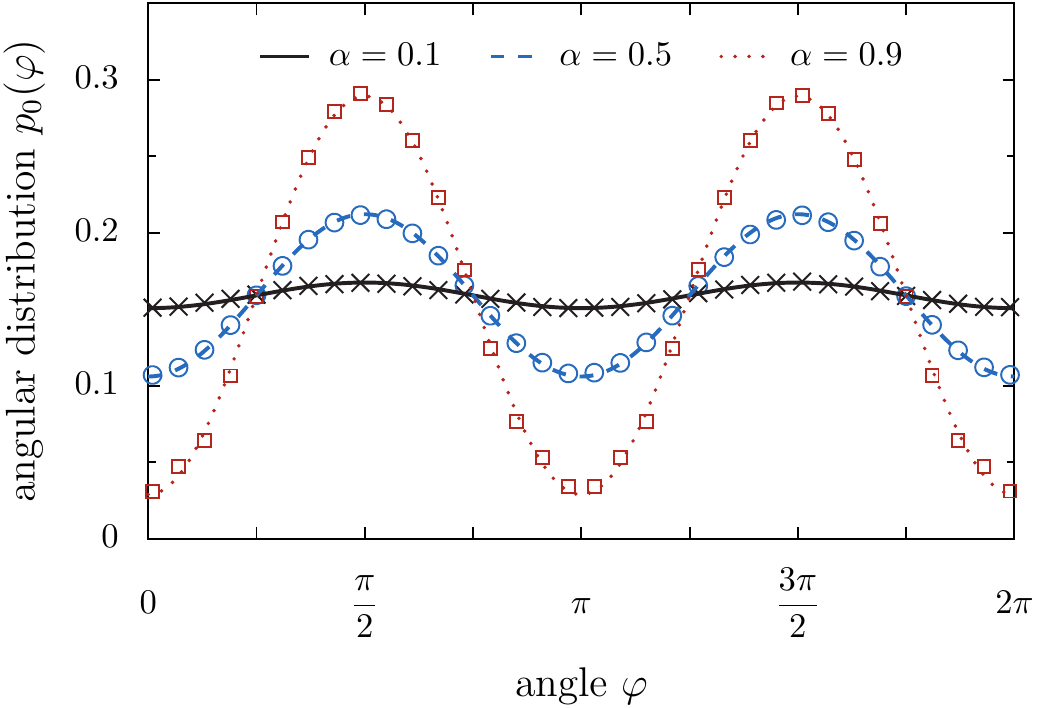}
\end{center}
\caption{Angular Distribution $p_0(\varphi)$ for different values of the anisotropy
$\alpha$. Lines represent theoretical results, cf. \eqref{eqn:exam:StatAngDist}, and points show results of numerical
Langevin simulations of the particle dynamics. Parameters in numerical simulation: $\#$particles $N=10\,000$, $\Delta
t = 4
\times 10^{-3}$, $D = 1$, $v_0 = 1$. }
\label{fig:AngDist:aniso}
\end{figure}

These results illustrate our general discussion in section \ref{sec:GenMod}. SPP motion, as it is considered
in this work, is a combination of two processes: (i)  motion at constant speed along the director and (ii)
Brownian motion of an imaginary particle on a manifold described by the director. In turn, the imaginary particle is found 
with equal
probability on every
point of the manifold in the long-time limit since the dynamics of this particle is Brownian. This is reflected
by expression \eqref{eqn:stat:prob:ani:l}. However, the distribution of the direction of motion $\varphi$ may be
anisotropic as indicated by expression 
\eqref{eqn:stat:prob:ani:phi} depending on the way the manifold is
embedded in space. The central quantity which relates the motion of the imaginary particle on the manifold to the
direction of motion of the SPP is the metric tensor $g(\varphi)$. 

The general solution given by equation \eqref{eqn:gen:sol:2d:P:ani:11} allows us to estimate dynamic quantities
as well, such as the
characteristic correlation time. For large observation times $t \gg t_0$, the
relaxation of the angular probability distribution is determined by the slowest mode: 
\begin{align} 
 \label{eqn:gen:sol:2d:P:ani:12}
 p(\varphi,t| \varphi_0,t_0) \simeq \frac{\sqrt{g(\varphi)}}{L} \, \left \{ 1+2 \cos \left [
\frac{2 \pi}{L} \int_{\varphi_0}^\varphi \! \d \varphi' \, \sqrt{g(\varphi')} \,\right ] e^{- \left
(2 \pi/L \right )^{2} D \, (t-t_0) } \right \} \! .
\end{align}
Thus, the characteristic relaxation time reads
\begin{align} 
  \tau_{c} \approx \frac{L^2}{4 \pi^2 D}. 
\end{align}
The perimeter of a circle equals $L=2\pi$ confirming the prediction of isotropic self-propelled motion in two
dimensions, cf. \eqref{eqn:corr:func:time:2d}: $\tau_c = D^{-1}$. Note, however, that for a deformed circle whose
perimeter is kept
constant at $L=2\pi$, the overall relaxation time does not change but is still determined by 
\begin{align} 
  \label{eqn:gen:iso:corr:time:21}
  \tau_{c} \approx D^{-1}. 
\end{align}
Thus, the correlation function is expected to decay exponentially with a characteristic timescale $\tau_{c}$, when
initial transients are neglected. The length of the transient depends on the timescale separation between the first and
the second mode in \eqref{eqn:gen:sol:2d:P:ani:11}. Interestingly, the ratio of characteristic times of the
leading and the next-to leading order mode does not depend on model parameters but is a fixed value: 
\begin{align*} 
   \left ( \frac{L}{2 \pi \sqrt{D}} \right )^2 \!\left / \left ( \frac{L}{4 \pi \sqrt{D}} \right )^2 \right. =
4.
\end{align*}
This time scale separation is large enough to be observed experimentally or in simulations and justifies our reasoning
above. The estimation of the correlation time allows us to evaluate the diffusion coefficient via the general relation
\eqref{eqn:gen:diff:coeff:kobu:21}: 
 \begin{align}
  \label{eqn:apprxo:diff:coeff:exm:8}
  \mathcal{D}_x = \left ( \frac{L}{2\pi} \right )^{\!2} \cdot \frac{v_0^2}{2 D}.   
 \end{align}
Interestingly, the  diffusion coefficient does not depend on the degree of anisotropy. 

\begin{figure}[tb]
\begin{center}
  \includegraphics[width=\textwidth]{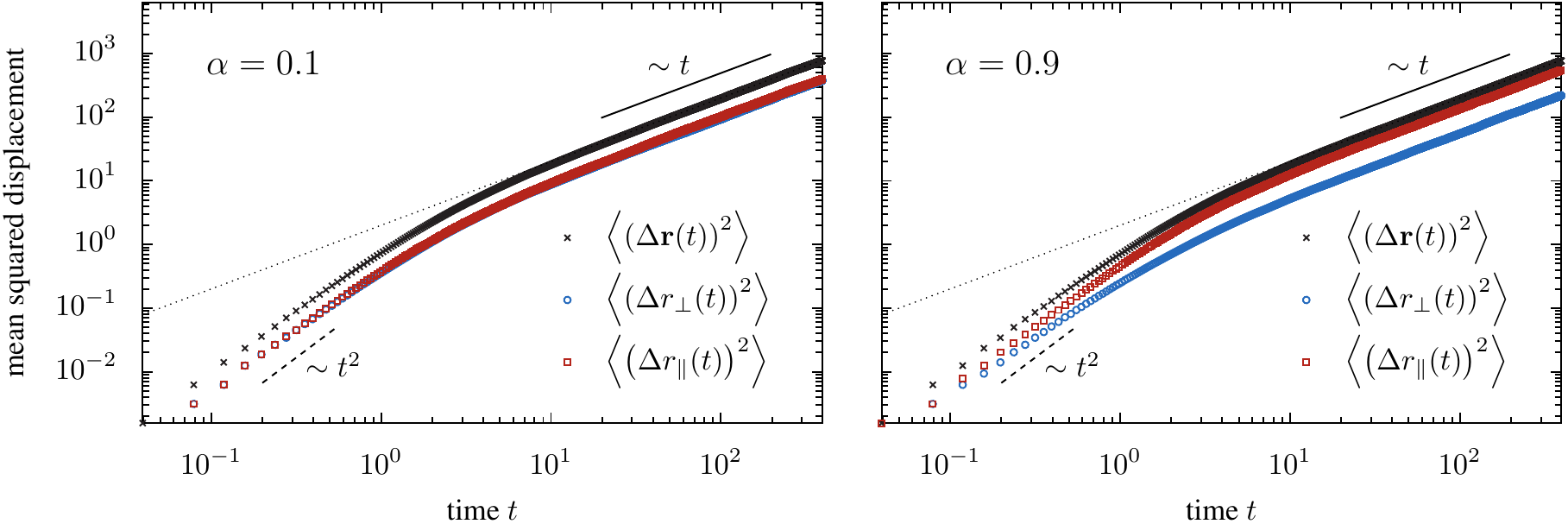}
\end{center}
\caption{Mean squared displacement as a function of time for two different anisotropies: $\alpha = 0.1$ (left) and
$\alpha = 0.9$ (right). The full mean squared displacement as well as the components parallel and perpendicular to the
preferred axis are shown. The dotted line indicates the limiting behavior $\mean{\left ( \Delta \vec{r}(t) \right )^2}
\sim 4 \mathcal{D}_x t$ with the approximate diffusion coefficient $\mathcal{D}_x = \tau_{c} v_0^2/2$, cf.
\eqref{eqn:apprxo:diff:coeff:exm:8}. Parameters in numerical simulation: $\#$particles $N=10\,000$, $\Delta t = 4 \times
10^{-3}$, $D = 1$,
$v_0 = 1$. }
\label{fig:MSD:aniso}
\end{figure}

In order to visualize our results, we consider an illustrative example where the metric is given by
 \begin{align}
  \label{eqn:exam:metric}
      g(\varphi) = \left [ \frac{1 - \alpha \cos^2 \varphi}{1-\alpha/2}  \right ]^2.  
 \end{align}
The degree of anisotropy is measured by the parameter $\alpha \in [0,1]$. 
For $\alpha = 0$, the metric tensor is equal to a constant and isotropic motion is expected. 
% 
%In contrast, the metric tensor possesses maxima at $\bar{\varphi}_+ = \pi/2 + n \pi$ and is zero for $\bar{\varphi}_- =
% n \pi$ where $n \in \mathbb{Z}$. 
%
If $\alpha$ is larger than zero, the metric tensor possesses maxima at $\bar{\varphi}_+ = \pi/2 + n \pi$ and minima for $\bar{\varphi}_- = n \pi$ where $n \in \mathbb{Z}$. 
According to Eq.~(\ref{eqn:Langevin:Aniso:2d}), angular fluctuations are suppressed if the particle moves along the
$y$-axis of the laboratory frame, whereas fluctuations are increased if the particle moves parallel to the $x$-axis. 
Thus, the diffusion of the SPP is anisotropic. We denote the mean squared displacement along the
$x$-axis by $\mean{\left ( \Delta r_{\perp}(t) \right )^2}$ and along the $y$-axis by $\mean{\left ( \Delta
r_{\parallel}(t) \right )^2}$, where the indices $\perp$ and $\parallel$ indicate the mean squared
displacement perpendicular and parallel with respect to the preferred direction of motion ($y$-axis).

\begin{figure}[tb]
 \begin{center}
   \includegraphics[width=0.6\textwidth]{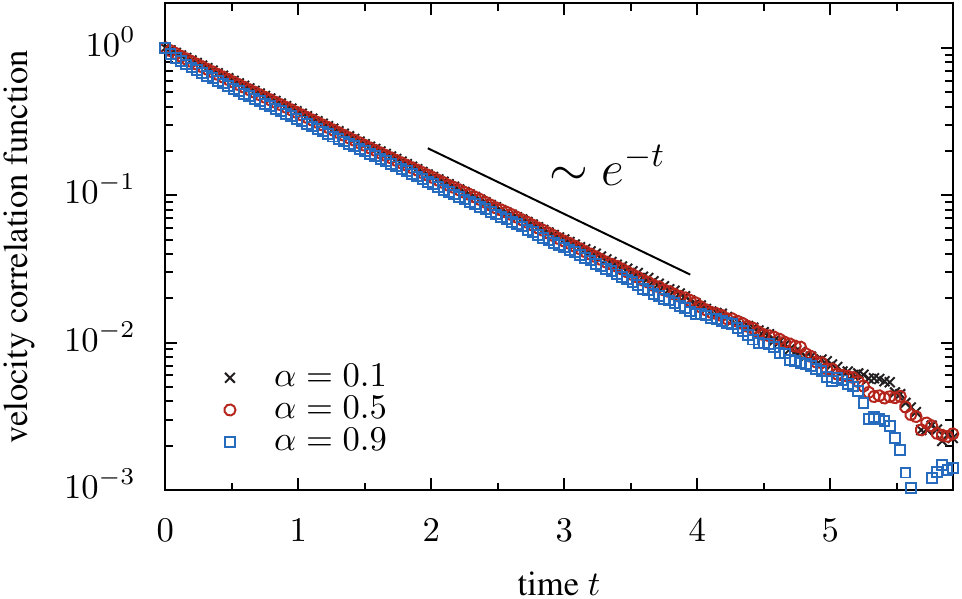}
 \end{center}
\caption{Velocity correlation function $\mean{\vec{v}(t) \cdot \vec{v}(0)}$ obtained from particle based Langevin
simulations of the microscopic dynamics \eqref{eqn:Langevin:Aniso:2d}. The correlation function is averaged over all
initial directions of motion. After a short initial transient, the velocity correlation decays exponentially with a
characteristic time which is determined by the slowest mode in the probability density function
\eqref{eqn:gen:sol:2d:P:ani:11}: $\tau_{c} = D^{-1}$. Parameters in numerical simulation: $\#$particles $N=500\,000$,
$\Delta t = 4
\times 10^{-4}$, $D = 1$, $v_0 = 1$. }
\label{fig:VCF:aniso}
\end{figure}

We have chosen this example (rather than the
ellipse shown in Fig. \ref{fig:ellipse}) because the resulting equations are compact and therefore easier to interpret.
In Fig. \ref{fig:AngDist:aniso}, we plot the stationary angular distribution
 \begin{align}
   \label{eqn:exam:StatAngDist}
   p_0(\varphi) = \frac{1 - \alpha \cos^2 \varphi}{(2-\alpha) \pi}
 \end{align}
for different anisotropies. The dependence of the arc length parameter $l$ on the polar angle reads in this case:
 \begin{align}
   l(\varphi) = \varphi - \frac{\alpha}{2-\alpha} \cdot \frac{\sin(2 \varphi)}{2} .
 \end{align}
Measurements of the mean squared displacement as well as the velocity correlation function from numerical Langevin
simulation of \eqref{eqn:Langevin:Aniso:2d} are depicted in Fig. \ref{fig:MSD:aniso} and Fig. \ref{fig:VCF:aniso},
respectively. Interestingly,
the long time behavior of the mean squared displacement does \textit{not} change due to the anisotropy in accordance
with \eqref{eqn:apprxo:diff:coeff:exm:8}. However, one can observe a splitting of the parallel $\mean{\abs{\Delta
r_{\parallel}}^2}$ and perpendicular $\mean{\abs{\Delta r_{\perp}}^2}$ component of the mean squared displacement with
increasing anisotropies, see Fig. \ref{fig:MSD:aniso}. Moreover, the predictions on the long-time behavior of the
correlation function and the characteristic correlation time agree well with the prediction
\eqref{eqn:gen:iso:corr:time:21}, cf. Fig. \ref{fig:VCF:aniso}: the correlation function does possess
exponential tails
with a characteristic time determined by $\tau_c = D^{-1}$. 

\section{Summary \& Outlook}
\label{sec:Summary}

We presented a geometric view onto SPP motion at constant speed. 
In order to do so, we divided the problem into two separate parts: (i) motion at constant speed in the direction
given by the director in the $d$-dimensional physical space; (ii) stochastic dynamics of the director itself which
takes place on a curved, abstract, $(d-1)$-dimensional compact manifold. 
We saw that process (i) and (ii) are intimately linked. For example, the mean squared displacement
can immediately be obtained from the correlations of the director. 
This implies that all relevant information is contained in the dynamics of the director which can be mapped to
Brownian motion -- characterized by a noise strength $D$ -- on a compact manifold. 
One essential step of this approach has been the parametrization of the manifold using generalized
coordinates or angles. 
We showed that the dynamics of each generalized coordinate is described by a Langevin equation with a stochastic and a
deterministic term, both proportional to $D$. 
We used this formalism to indicate that SPPs perform isotropic
diffusion in the $d$-dimensional physical space if the manifold corresponds to a hypersphere retrieving
previous results~\cite{schienbein_langevin_1993,mikhailov_self_1997}. 
We also showed that the SPP performs anisotropic diffusion if the manifold does not correspond to a hypersphere, but
to the surface of an arbitrary surface embedded in $d$-dimensional physical space such that Eq.~\eqref{eq:manifold} is
obeyed.  
In summary, the proposed approach provides an unified framework to model isotropic as well as anisotropic diffusion of
SPPs in any dimension. 

A remark on the origin of anisotropic diffusion is in order here: 
The presence of a force field induces, asymptotically, ballistic motion, while we describe SPP with an
anisotropic directional persistence remaining asymptotically diffusive. 
These two problems -- anisotropic diffusion and random motion in a force field -- are fundamentally different. 
Anisotropic diffusion is often caused by an
anisotropic substrate. For example, the motion of eukaryotic cells moving on a pre-patterned surface exhibits a
preferred axis of motion~\cite{Csucs_locomotion_2007,ziebert_effects_2013}. 
A similar reasoning may apply to gliding bacteria where, for instance, trails of polysaccharides on the
substrate can induce anisotropic motion of the bacteria~\cite{dworkin}.   

Finally, the proposed geometric approach to SPP motion provides a simple framework to describe the motion of cells,
bacteria or other microorganisms in isotropic as well as anisotropic environments. It may be viewed as a coarse-grained
description of the actual system under consideration in the sense that a complex self-propelled object moving on an
anisotropic substrate is modeled by a point particle with anisotropic directional persistence. In this regard, it is
essential that model parameters are immediately linked to observables which
are easily accessible experimentally. We have illustrated this statement by a two dimensional example studied in
section \ref{sec:AnIsoSpp}. The stationary probability distribution to find a particle moving in a certain direction
sheds light on the anisotropy of angular fluctuations. Similarly, the amplitude of these fluctuations enters into both
the diffusion coefficient and the characteristic relaxation time of the velocity correlation function. Once these
quantities have been measured, the point particle model can caricature the motion of a real-world active particle
sufficiently well. 

%%%%%%%%%%%%%%%%%%%%%%%%%%%%%%%%%%%%%%%%%%%%%%%%%%%%%%%%%%%%%%%%%%%%%%%%%%%%%%%%%%%%%%%%%%%%%%%%%%%%%%%%%%%
\appendix

\section{Diffusion on n-spheres}
\label{sec:App:DDDDD}

For completeness, we state the equations of motion of a Brownian particle diffusing on a $n$-sphere. This
surface, embedded in $d=n+1$ dimensions, can be parametrized by $n$ angles. We use the convention
\cite{blumenson_derivation_1960}
 \begin{align*}
  \begin{split}
     e_1    &= \cos \varphi_1, \\
     e_{\mu}    &= \cos \varphi_{\mu} \prod_{m=1}^{\mu-1} \sin \varphi_m, \quad \mu=2,...,d-1\\
     e_{d}  &= \prod_{m=1}^{d-1} \sin \varphi_m,
  \end{split}
 \end{align*}
where $\varphi_{1,...,d-2} \in [0,\pi)$ and $\varphi_{d-1} \in [0,2\pi)$. The metric of the sphere in this coordinates
is diagonal and reads
 \begin{align*}
  g_{\mu \nu} = \delta_{\mu \nu} \begin{cases}
                                   1 & \mu = 1, \\ 
				    \prod_{m=1}^{\mu-1} \sin^2 \varphi_m & \mu = 2,3,...,d-1. 
                                 \end{cases}
 \end{align*} 
Thus, we obtain the following Langevin equations for the angles \cite{zinn_quantum_2002}:
\begin{subequations}
 \begin{align*}
  \mbox{(S)} \;\;\quad \frac{\d \varphi_1}{\d t}    \;\,  &= \sqrt{2 D} \, \xi_1(t) + D \frac{d-2}{\tan \varphi_1}, \\
  \mbox{(S)} \;\;\quad \frac{\d \varphi_\mu}{\d t}  \;\,  &= \frac{\sqrt{2 D}}{\prod_{m=1}^{\mu-1} \sin \varphi_m} \,
  \xi_\mu(t) + D \frac{d-\mu-1}{\tan \varphi_\mu \prod_{m=1}^{\mu - 1} \sin^2 \varphi_m} , \quad \mu=2,3,...,d-2,  \\
  \mbox{(S)} \quad \frac{\d \varphi_{d-1}}{\d t}  &= \frac{\sqrt{2 D}}{\prod_{m=1}^{d-2} \sin \varphi_m} \, \xi_{d-1}(t)
.
 \end{align*}  
\end{subequations}
This set of equations in combination with Eq. \eqref{eq:first} are convenient to simulate self-propelled motion in
$d$ spatial dimensions. 
% 

%%%%%%%%%%%%%%%%%%%%%%%%%%%%%%%%%%%%%%%%%%%%%%%%%%%%%%%%%%%%%%%%%%%%%%%%%%%%%%%%%%%%%%%%%%%%%%%%%%%%%%%%%%%
% refs

% \bibliographystyle{MyEPJST}
% \bibliography{Anistropic}

% \section*{References}

\end{document}